\documentclass[aip, amsmath, amssymb, reprint]{revtex4-1}

\usepackage{graphicx}
\usepackage{bm}
\usepackage[T1]{fontenc}
\usepackage[utf8]{inputenc}
\usepackage{xurl}
\usepackage{esint}

\usepackage[unicode=true, bookmarks=true, bookmarksnumbered=true, bookmarksopen=true, bookmarksopenlevel=1, breaklinks=false, pdfborder={0 0 0}, backref=false, colorlinks=true]{hyperref}
\hypersetup{citecolor=blue, linkcolor=blue, urlcolor=blue}

\usepackage[dvipsnames]{xcolor}

\usepackage[english]{babel}
\usepackage[autostyle, english = american]{csquotes}
\MakeOuterQuote{"}

\usepackage{mathptmx}
\usepackage{etoolbox}

\makeatletter
\def\@email#1#2{%
 \endgroup
 \patchcmd{\titleblock@produce}
  {\frontmatter@RRAPformat}
  {\frontmatter@RRAPformat{\produce@RRAP{*#1\href{mailto:#2}{#2}}}\frontmatter@RRAPformat}
  {}{}
}%
\makeatother

\begin{document}

\title{Coordinate-invariant flux-surface Fourier analysis in tokamaks}

\author{M. Pharr$^*$}\email{mcp2198@columbia.edu}
\author{E. Bursch}
\author{N.C. Logan}
\author{P. Lunia}
\affiliation{Department of Applied Physics and Applied Mathematics, Columbia University, New York, NY 10027, USA}
\affiliation{Columbia Fusion Research Center, Columbia University, New York, NY 10027, USA}

\author{J.K. Park}
\affiliation{Department of Nuclear Engineering, Seoul National University, Seoul 08826, Republic of Korea}

\author{C. Paz-Soldan}
\affiliation{Department of Applied Physics and Applied Mathematics, Columbia University, New York, NY 10027, USA}
\affiliation{Columbia Fusion Research Center, Columbia University, New York, NY 10027, USA}

\date{\today}

\begin{abstract}
The Fourier spectra of resonant quantities in tokamaks depend on the choice of magnetic coordinates, and an area weighting of the Fourier integrand preserves the resonant coefficients on rational surfaces\cite{parkSpectralAsymmetryDue2008}. That result constrains only the resonant interior; the coordinate dependence of the external Fourier spectrum, which determines the coupling to Resonant Magnetic Perturbation (RMP) coils and error-field penetration, was left untreated. This paper shows that pairing a square-root-area weighted vacuum field perturbation with a full-area-weighted resonant field yields a coupling matrix $C$ whose singular values are invariant under coordinate transformations and whose right singular vectors reconstruct to a consistent real-space field pattern across coordinate systems, completing the coordinate-invariance picture for the plasma-3D-field coupling paradigm. GPEC calculations confirm the analytic result and show that improperly weighted coupling matrices can produce dominant modes whose overlap with the vacuum field perturbation differs by a factor of 2--3 between coordinate systems for strongly shaped, low aspect ratio equilibria, with the discrepancy growing with inverse aspect ratio. The same coordinate dependence afflicts alternative formulations such as the three-mode metric or zeroing the $q=2$ resonant field without proper weighting. The result applies to any tool computing Fourier spectra of resonant or external quantities on flux surfaces.
\end{abstract}

\maketitle

\section{\label{sec:intro} Introduction}

Resonant coupling is the paradigm of plasma-3D-field coupling in which the coupling between an external magnetic field perturbation and the plasma is quantified by a coupling matrix $C$ that relates a resonant metric at rational surfaces, such as the resonant flux or the resonant normal field, to the Fourier spectrum of the vacuum field perturbation at the control surface. The control surface is the flux surface, typically the plasma boundary, on which the external field perturbation is decomposed to construct $C$; this terminology and the underlying coupling-matrix formulation were established in foundational work on resonant coupling\cite{parkComputationThreedimensionalTokamak2007,parkControlAsymmetricMagnetic2007,parkErrorFieldCorrection2008,parkIdealPerturbedEquilibria2010}. This paradigm has been used in a number of works to design RMP coils for both edge-localized mode (ELM) suppression and error-field correction (EFC)\cite{loganSPARCTokamakError2026,leutholdARCPhysicsBasis2026,pigattoModelingdrivenRequirementsError2026a,loganPhysicsBasisDesign2021,sweeneyMHDStabilityDisruptions2020,yangPermanentMagnetsELM2026}, to assess error fields in tokamaks \cite{loganSPARCTokamakError2026,pharrErrorFieldPredictability2024,baiTimeVariationError2025,burschImprovedN1Empirical2026,loganEmpiricalScaling22020,loganRobustnessTokamakError2020,amoskovAssessment1Overlap2019,parkMDC19ReportAssessment2017,pigattoModelingdrivenRequirementsError2026a,parkErrorFieldCorrection2008,butteryLimitsChallengesError2012}, to model comprehensive 3D field strategies\cite{pharrQuantifyingResonantDrive2026a,yangLocalizingResonantMagnetic2020,yangTailoringTokamakError2024,parkQuasisymmetricCorrectionNonresonant2025,parkQuasisymmetricOptimizationNonaxisymmetry2021,peterkaStudyMHDStability2026,loganMetricsExtrapolationResonant2025}, and to model experimental 3D field results\cite{paz-soldanImportanceMatchedPoloidal2014,parkIdealPerturbedEquilibria2010,paz-soldanSpectralBasisOptimal2014,parkErrorFieldCorrection2011,luniaPredictedThresholdsRMP2025,igochinePlasmaEffectError2023,yangParametricDependenciesLocked2021,yangPermanentMagnetsELM2026}. The widespread use of this paradigm motivates ensuring that the results obtained from it are robust to coordinate system choice.

The Fourier spectra of resonant quantities in tokamaks depend on the choice of magnetic coordinates, and a Fourier integrand weighting of the differential area element preserves the resonant Fourier coefficients of quantities at rational surfaces\cite{parkSpectralAsymmetryDue2008}. This allows one to construct a resonant field vector that consists of the $m$th area-weighted field component on the $q_{ith} = m/n$ surface, which is entirely invariant to coordinate system choice.

However, this only addresses resonant coefficients in the interior of the plasma, and therefore the co-domain of a resonant coupling matrix. The coordinate dependence of the full external Fourier spectrum, which determines the coupling to RMP coils and error-field penetration, was not explicitly treated. Coordinate system-dependent results remain common in the literature. The three-mode metric, also known as the three-mode error index (TMEI), is still in active use, with recent error-field studies on DTT\cite{albaneseErrorFieldCorrection2023} and JT-60\cite{koNumericalStudyNovel2025} adopting this framing, and other recent work\cite{sugiyamaApplicationNonintegerRMP2026} defines resonant metrics without explicit attention to the integrand weighting. Although GPEC is the most widely used implementation of the coupling-matrix paradigm, the result of this paper applies equally to other tools that compute Fourier spectra on flux surfaces, including recent MARS-F\cite{pigattoModelingdrivenRequirementsError2026a,mavigliaStudiesEUDEMO3D2023,liuModellingIntrinsicError2014} adoptions of the formalism.

This paper shows that both a square-root-area weighted vacuum perturbed field spectrum on the control surface and full-area-weighted resonant field components on rational surfaces are necessary to form a coupling matrix $C$ whose singular values are invariant under coordinate transformations, and whose right singular vectors have a consistent real-space structure across coordinate systems. This analytic result is confirmed with numerical calculations using GPEC\cite{parkIdealPerturbedEquilibria2009}, which both confirm the coordinate invariance of the SVD of $C$ when weighted with the square-root-area factor $\sqrt{\mathcal{J}\left|\nabla \psi\right|}$, and show that other weightings of the vacuum field perturbation lead to significant coordinate dependence of the SVD of $C$. We also present further confirmation of the results of Park 2008\cite{parkSpectralAsymmetryDue2008} that the resonant Fourier coefficients of the full-area-weighted resonant field are invariant under coordinate transformations, while those of the square-root-area weighted and bare resonant fields are not. 

This paper is organized as follows: section \ref{sec:coord_invariance} presents the analytic proof of the coordinate invariance of the SVD of $C$ when square-root-area weighted. Section \ref{sec:overlap} presents the dominant mode overlap as a coordinate-invariant metric of resonant coupling, and clears up confusion in past literature about the units and definition of $\delta$. Section \ref{sec:discussion} discusses the implications of these results for past literature and for future work.

\section{\label{sec:coord_invariance} Coordinate Invariance of Resonant Coupling}

\begin{figure}
\includegraphics[width=0.8\linewidth]{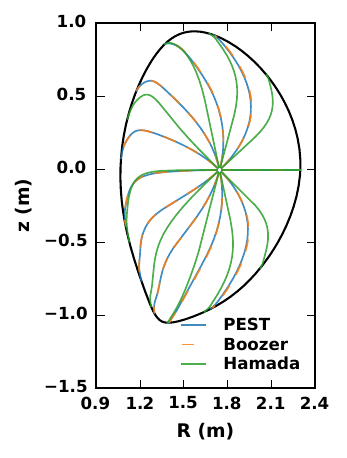}\caption{Constant $\theta$ lines shown at zero machine angle in PEST, Boozer, and Hamada coordinates for a DIII-D-like equilibrium used in previous works\cite{burgessTearingStabilityPrediction2026,pharrQuantifyingResonantDrive2026a}. Note that PEST and Boozer coordinates are similar, while Hamada coordinates show significant deviation, especially a tendency to group near the x-point.\label{fig:coords}}
\end{figure}

\begin{figure*}
\includegraphics[width=\linewidth]{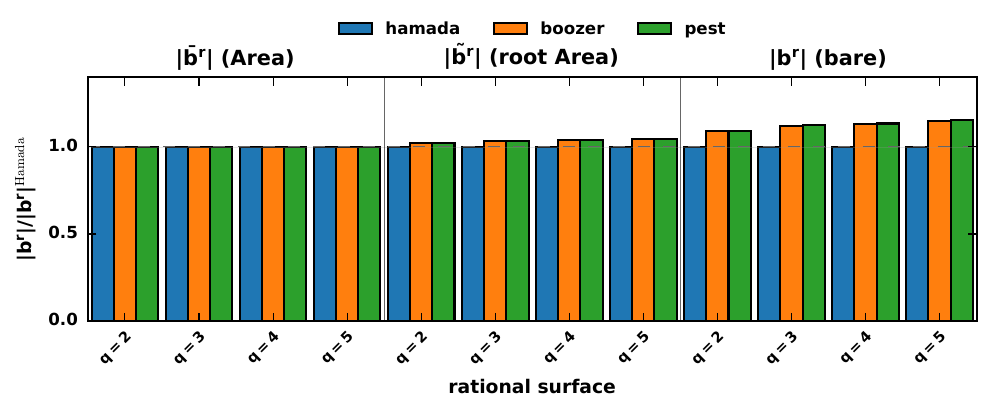}\caption{Coordinate-invariance of the full-area-weighted resonant metric, $\bar{\mathbf{b}}^r$ (left), is contrasted with the coordinate dependence of the other weightings, ${\mathbf{b}}^r$ and $\tilde{\mathbf{b}}^r$.\label{fig:phires}}
\end{figure*}

\begin{figure}
\includegraphics[width=\linewidth,height=0.9\textheight,keepaspectratio]{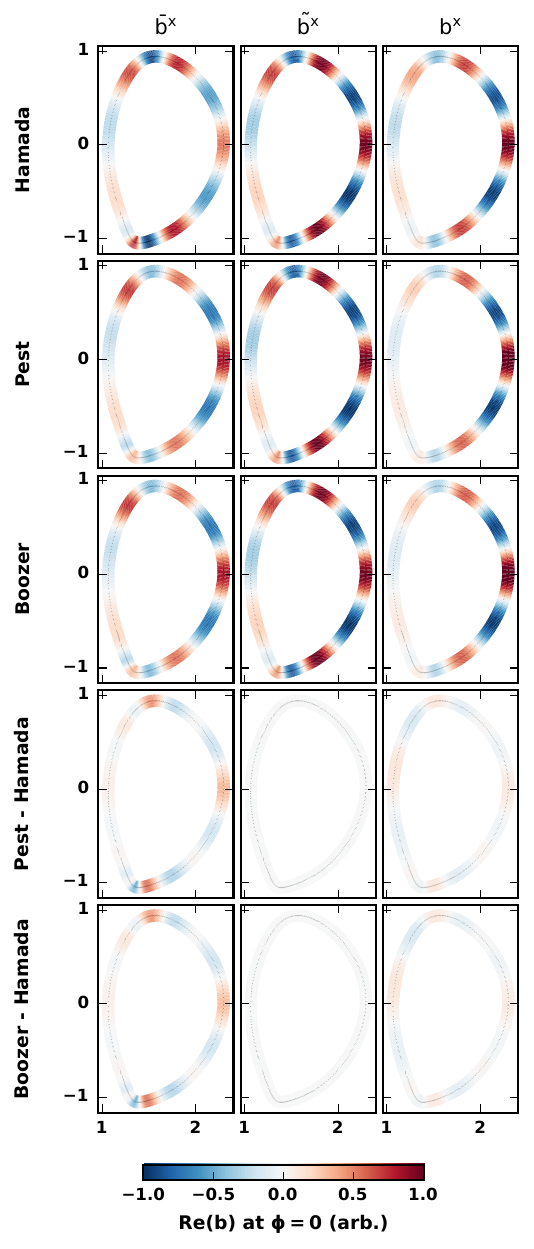}\caption{First right-singular vector, i.e. the dominant mode, of the coupling matrix $C$ defined on the control surface of an equilibrium cross section for the different weightings of the vacuum field perturbation shown in real space. Above are the dominant modes calculated using three different coordinate systems, and below are the differences between them. Only the dominant mode constructed with $\tilde{\mathbf{b}}^x$ is identical when calculated using different magnetic coordinates, shown by the differences in the bottom center panels being negligible.\label{fig:singvecs}}
\end{figure}

In the presence of closed flux surfaces, an ideal MHD equilibrium magnetic field in a tokamak can be expressed in straight field line magnetic coordinates $(\psi, \theta, \phi)$ as,
\begin{equation}
\mathbf{B} = B_0 \left(\nabla \phi \times \nabla \psi + q\left(\psi\right)\nabla \psi \times \nabla \theta\right). \label{eq:contra}
\end{equation}
In Eq.~\eqref{eq:contra}, the units of each quantity are dependent on the choice of magnetic coordinates. $\psi$ must simply label flux surfaces; it can be normalized to be dimensionless or have units, and can be defined in a number of ways, such as a poloidal or a toroidal flux, a pressure, a radius, or a volume. $\theta$ and $\phi$ must be periodic coordinates that label points on a flux surface, but they can be defined in a number of ways, for example as geometric angles, specific straight-field-line angles, Boozer angles\cite{boozerPlasmaEquilibriumRational1981,boozerEstablishmentMagneticCoordinates1982}, or Hamada angles. $B_0$ is a constant with or without units, whose dimensions and magnitude depend on on the units and normalization of $\psi$. $q$ is the safety factor, which is a function of $\psi$ and is dimensionless. The Jacobian of the coordinate system is given by
\begin{equation}
    \mathcal{J} = \left(\nabla \psi \cdot \nabla \theta \times \nabla \phi\right)^{-1} = B_0 \left(\mathbf{B} \cdot \nabla \theta\right)^{-1} = B_0 q \left(\mathbf{B}\cdot\nabla\phi\right)^{-1}, \label{eq:jacobian}
\end{equation}
meaning that the units of the Jacobian are also dependent on the choice of magnetic coordinates. Importantly, the differing choice of magnetic coordinates, particularly the poloidal angle, leads to poloidal Fourier spectra of various physical quantities that are different in different coordinate systems. For a general overview of this topic, see D'haeseleer et al\cite{dhaeseleerFluxCoordinatesMagnetic1991}.

Magnetic coordinates in general have a two-degree gauge freedom, and different schemes exist to constrain this gauge freedom, which defines a coordinate system. In straight-field-line coordinate systems, one of these degrees of freedom is used to ensure magnetic field lines are straight; the remaining degree of freedom allows angles to vary with any function of space as\cite{parkSpectralAsymmetryDue2008,greeneStabilityCriterionArbitrary1962},
\begin{align}
    \theta &\rightarrow \theta + f\left(\mathbf{x}\right)/q\label{eq:trans1},\\
    \phi &\rightarrow \phi + f(\mathbf{x}).\label{eq:trans2}
\end{align} 
This gauge freedom leads to phase changing whenever $f$ is a function of $\psi$ alone, and to Fourier mode mixing whenever $f$ is a function of $\theta$ and/or $\phi$. It is this reason why Fourier spectra of various physical quantities differ between coordinate systems, as illustrated for the three coordinate systems used throughout this paper in Fig.~\ref{fig:coords}.

For the vacuum field perturbation at the control surface, three natural Fourier decompositions exist for a given toroidal mode number, distinguished by the area weight applied to the integrand. We work in field units (Tesla) throughout, dividing each integral by the appropriate power of the coordinate-invariant control surface area $A^x = \oiint dA$:
\begin{align}
    b^x_{mn} &= \oiint d\theta \, d\phi \,\, \delta \mathbf{B}^x \cdot \hat n \, e^{i(n\phi-m\theta)}, \label{eq:b_bare}\\
    \tilde b^x_{mn} &= \frac{1}{\sqrt{A^x}}\oiint d\theta \, d\phi \,\, \delta \mathbf{B}^x \cdot \hat n \, \sqrt{\mathcal{J} |\nabla \psi|} \, e^{i(n\phi-m\theta)}, \label{eq:b_half}\\
    \bar{b}^x_{mn} &= \frac{1}{A^x}\oiint d\theta \, d\phi \,\, \delta \mathbf{B}^x \cdot \hat n \, \mathcal{J} |\nabla \psi| \, e^{i(n\phi-m\theta)}. \label{eq:b_full}
\end{align}
\noindent Here $\delta \mathbf{B}^x$ is the vacuum field perturbation, the perturbed magnetic field not including the plasma response. The $x$ superscript denotes `external'. The decoration encodes the Jacobian weight in the integrand: $b^x$ has none, $\tilde b^x$ has one factor of $\sqrt{\mathcal{J}|\nabla\psi|}$, and $\bar b^x$ has the full Jacobian $\mathcal{J}|\nabla\psi| = dA/(d\theta\, d\phi)$. The matching power of $A^x$ in the prefactor makes all three quantities carry units of Tesla, and is what makes the coupling matrix $C$ defined below dimensionless and properly weighted.

Introducing the gauge transformation in Eqs.~\eqref{eq:trans1} and \eqref{eq:trans2} produces a factor of $f(\mathbf{x})$ in the Fourier exponent together with factors of $1+\partial f/\partial \phi$ and $1+\partial(f/q)/\partial \theta$ in the integrand. For $\mathbf{b}^x$ and $\bar{\mathbf{b}}^x$, this is neither an identical nor a unitary operation, so their Fourier spectra do not transform consistently across coordinate systems. The square-root-area weighted form $\tilde{\mathbf{b}}^x$ has a special property. Parseval's theorem gives the $L^2$ norm of $\tilde{\mathbf{b}}^x$ as
\begin{multline}
    \tilde{\mathbf{b}}^{x\dagger} \tilde{\mathbf{b}}^x = \sum_{m,n} \tilde{b}^{x*}_{mn} \, \tilde{b}^x_{mn} = \frac{1}{A^x}\oiint d\theta\, d\phi \,\, (\delta \mathbf{B}^x \cdot \hat{n})^2 \mathcal{J} |\nabla \psi| \\ = \frac{1}{A^x}\oiint dA \, (\delta \mathbf{B}^x \cdot \hat{n})^2 = \langle (\delta \mathbf{B}^x \cdot \hat{n})^2 \rangle,
\end{multline}
the area-averaged squared normal perturbed field at the control surface. This is a physical quantity that must be coordinate invariant. In contrast, the $L^2$ norms of $\mathbf{b}^x$ and $\bar{\mathbf{b}}^x$ are not invariant and can be made arbitrarily small or large by choosing a coordinate system with a small or large Jacobian, respectively.

The converse identifies the square-root-area weighting as the unique admissible weighting for the resulting Fourier coefficients to transform unitarily under coordinate transformations. Consider the Fourier decomposition of the normal field with an arbitrary integrand weight $W(\theta,\phi)$,
\begin{equation*}
    b^W_{mn} = \oiint d\theta \, d\phi \,\, \delta \mathbf{B}^x \cdot \hat{n} \, W \, e^{i(n\phi-m\theta)}.
\end{equation*}
By Parseval's theorem, its $L^2$ norm is $\sum_{m,n} |b^W_{mn}|^2 = \oiint d\theta \, d\phi \,\, (\delta \mathbf{B}^x \cdot \hat{n})^2 \, W^2$. The coefficient vector transforms unitarily between coordinate systems if and only if this norm is invariant for an arbitrary normal field, which in turn holds precisely when $W^2 \, d\theta \, d\phi$ is a coordinate-invariant measure on the control surface. This is due to the fact that $\mathcal{J}\left|\nabla\psi\right|$ is the Jacobian of the transformation from the coordinates $\theta ,\, \phi$ to the surface in 3D space, so $\mathcal{J}\left|\nabla \psi\right|\,d\theta\,d\phi$ is the canonical measure of the surface which forms a Riemannian manifold\cite{kreyszigIntroductionDifferentialGeometry1968}. The physical area element $dA = \mathcal{J} |\nabla \psi| \, d\theta \, d\phi$ is, up to a coordinate-invariant scalar field, a unique such measure. This multiplicative scalar field freedom allows this measure to work for quantities other than the normal field, such as a current or displacement quantity, so long as that quantity is defined as a function of real space $\mathbf{x}$. The square-root weighting of Eq.~\eqref{eq:b_half} is therefore necessary, not only sufficient. The bare and full weightings fail this criterion because $W^2 = 1$ and $W^2 = (\mathcal{J} |\nabla \psi|)^2$ are not proportional to $\mathcal{J} |\nabla \psi|$.

The invariance of $\tilde{\mathbf{b}}^{x\dagger}\tilde{\mathbf{b}}^x$ implies that $\tilde{\mathbf{b}}^x$ is a vector quantity that transforms as
\begin{equation}
    \tilde{b}^{x\prime}_{m'n} = T_{m'm} \, \tilde{b}^x_{mn} \label{eq:trans3}
\end{equation}
\noindent where $T$ is a unitary coordinate transformation operator that can be constructed by finding the Fourier spectrum of each Fourier mode in the original coordinate system in the new coordinate system. No transformation for the toroidal coordinate is needed as toroidal modes change by at most a phase factor that can be zeroed by a suitable choice of zero angle. From here, we will assume a given toroidal mode number and only label the poloidal mode number in $\tilde{b}^x$. In contrast, $\mathbf{b}^x$ and $\bar{\mathbf{b}}^x$ do not transform consistently under these coordinate transformations.

On each rational surface $i$, we define the pitch-resonant normal field Fourier amplitude $\bar{b}^r_i = \bar{\Phi}^r_i / A^r_i$, where $\bar{\Phi}^r_i$ is the full-area-weighted Fourier component\cite{parkSpectralAsymmetryDue2008} that is resonant at rational surface $i$ and $A^r_i$ is the local area of that rational surface, not the control surface area $A^x$. Both normalizations are valid as far as coordinate invariance and units, but normalizing by $A^x$ would misweight the coupling matrix and disproportionately prioritize the high-$q$ outer surfaces. Park 2008 showed that on each rational surface $i$, the single Fourier coefficient resonant with that surface, specifically the $m = nq_i$ component evaluated locally about the surface, is invariant under coordinate transformations, though the rest of the poloidal spectrum on the same surface is not. Since $A^r_i$ is also coordinate invariant, the scalar $\bar{b}^r_i$ is invariant. The vector $\bar{\mathbf{b}}^r$ used below is therefore indexed by rational surface, not by poloidal mode number; its $i$th entry contains only the pitch-resonant amplitude on surface $i$.

Analogous square-root- and bare-weighted resonant amplitudes $\tilde{\mathbf{b}}^r$ and $\mathbf{b}^r$ can be defined by inserting one or no factors of $\sqrt{\mathcal{J}|\nabla\psi|}$ into the resonant Fourier integral, but only $\bar{\mathbf{b}}^r$ displays the invariance demonstrated in Park 2008\cite{parkSpectralAsymmetryDue2008}. Fig.~\ref{fig:phires} confirms this on a DIII-D-like equilibrium: across PEST, Boozer, and Hamada the per-surface amplitudes $|\bar{b}^r_i|$ agree to roundoff error, while $|\tilde{b}^r_i|$ and $|b^r_i|$ vary appreciably between coordinate systems. This $622$ kA, $-0.63$ T, pedestal-free equilibrium has been used in other works to refer to DIII-D-like scenarios\cite{burgessTearingStabilityPrediction2026,pharrQuantifyingResonantDrive2026a}. Note, as ideal GPEC has no resistivity, there is no resonant flux on the rational surface; however, plotted here and throughout this paper will be the {\it effective resonant flux} which shares the weighting invariance property of the actual resonant flux in resistive MHD. This is the amount of flux that is shielded by the singular current on the rational surface in ideal MHD. The derivation showing the definition of the effective flux in GPEC can be found in older works\cite{parkIdealPerturbedEquilibria2009}, and discussion of the relationship between the real flux and the effective flux in more recent work\cite{pharrQuantifyingResonantDrive2026a}.

The linear MHD coupling between $\bar{\mathbf{b}}^r$ and $\tilde{\mathbf{b}}^x$ can be calculated by a perturbed equilibrium code, giving the dimensionless coupling matrix $C_{im}$ such that
\begin{equation}
    \bar{b}^r_{i} = C_{im} \, \tilde{b}_{m}^x. \label{eq:coupling}
\end{equation}
The singular value decomposition (SVD) of $C$ can be expressed as
\begin{equation}
    C_{im} = U_{ip} \, \Sigma_{pq} \, V^*_{qm} \label{eq:svd}
\end{equation}
where $U$ and $V$ are generally non-square matrices of unit magnitude left- and right-singular vectors, respectively, and $\Sigma$ is a diagonal matrix of singular values. Substituting the SVD of $C$ into the coupling equation in Eq.~\eqref{eq:coupling} gives
\begin{equation}
    \bar{b}^r_{i} = U_{ip} \, \Sigma_{pq} \, V^*_{qm} \, \tilde{b}_{m}^x. \label{eq:coupling_svd}
\end{equation}

Applying the coordinate transformation of Eq.~\eqref{eq:trans3} to the coupling equation gives
\begin{align}
    \bar{b}^r_{i} &= C_{im} \, T^*_{mm'} \, \tilde{b}_{m'}^{x\prime} \label{eq:coupling_trans}\\
    &= U_{ip} \, \Sigma_{pq} \, V^*_{qm} \, T^*_{mm'} \, \tilde{b}_{m'}^{x\prime} \label{eq:coupling_svd_trans}\\
    &= U_{ip} \, \Sigma_{pq} \, \left(T_{m'm}V_{mq}\right)^\dagger \, \tilde{b}_{m'}^{x\prime}, \label{eq:coupling_svd_trans2}
\end{align}
where $*$ denotes complex conjugation and $\dagger$ denotes the conjugate transpose. Since the coefficients of $\bar{\mathbf{b}}^r$ are invariant, $\bar{b}^r_i = \bar{b}^{r\prime}_{i}$, and Eq.~\eqref{eq:coupling_svd_trans2} is itself an SVD of a coupling matrix $C'_{im'} = U_{ip}\Sigma_{pq}V'^{*}_{qm'}$ with $V' = TV$. Because $T$ is unitary, $V'$ is a valid matrix of right singular vectors. The SVD of $C$ is therefore coordinate-invariant, with the right singular vectors transforming as the Fourier spectra of the square-root-area weighted vacuum field perturbation. The same construction with $\mathbf{b}^x$ or $\bar{\mathbf{b}}^x$ in place of $\tilde{\mathbf{b}}^x$ obeys no corresponding unitary transformation, and the SVD of $C$ becomes coordinate-dependent.

\begin{figure*}
\includegraphics[width=\linewidth]{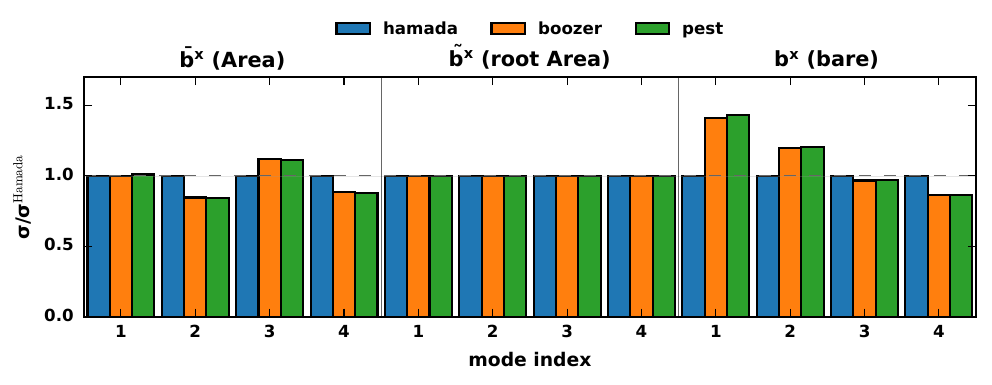}\caption{Singular values of the coupling matrix $C$ defined for the three different weightings of the vacuum field perturbation and an area-weighted resonant field calculated for $n=1$ in a DIII-D-like equilibrium shown in Fig.~\ref{fig:coords}. Singular values are shown normalized by their value in Hamada coordinates; $x$-axis `mode index' indicates the $i$th singular value ordered maximum to minimum. The coupling matrix constructed with $\tilde{\mathbf{b}}^x$ (middle) shows consistent singular values across different coordinate systems, while those with one more or less factor of $\sqrt{\mathcal{J}\left|\nabla \psi\right|}$ do not.\label{fig:singvals}}
\end{figure*}

The invariance of the singular values is shown in Fig.~\ref{fig:singvals}. The singular values of $C$ built from $\tilde{\mathbf{b}}^x$ agree across PEST, Boozer, and Hamada coordinates to numerical noise, while those built from other weightings of the vacuum field perturbation differ noticeably between coordinate systems. The first right singular vectors (RSVs) are reconstructed as physical fields on the control surface via inverse Fourier transform in Fig.~\ref{fig:singvecs}. The RSVs built from $\tilde{\mathbf{b}}^x$ are the same field patterns in all three coordinate systems, with no differences on the visualization scale, while the RSVs built from $\mathbf{b}^x$ or $\bar{\mathbf{b}}^x$ see peak local differences of $\sim$10\% of the maximum field value between coordinate systems.

Because $V' = T V$ transforms identically to $\tilde{\mathbf{b}}^{x\prime} = T \tilde{\mathbf{b}}^x$, each right singular vector of $C$ reconstructs to a unique physical field distribution in real space, independent of the magnetic coordinate system used to compute it. The right singular vectors are therefore a physical quantity, known as the ``dominant modes''\cite{parkErrorFieldCorrection2008,parkErrorFieldCorrection2011,paz-soldanImportanceMatchedPoloidal2014,loganIdentificationMultimodalPlasma2016,loganEmpiricalScaling22020,loganRobustnessTokamakError2020,loganPhysicsBasisDesign2021,yangLocalizingResonantMagnetic2020} in the plasma-3D-field coupling paradigm, and can be applied to plasma control problems. Those from improperly weighted coupling matrices are not, and can differ significantly enough to alter applications such as coil design and error-field correction.

\section{\label{sec:overlap} Dominant Mode Overlap}
One common application of this coordinate system invariant formulation is quantifying the effect of a given external magnetic field perturbation on tearing modes in a tokamak plasma \cite{bandyopadhyayMHDDisruptionsControl2025,loganEmpiricalScaling22020,loganRobustnessTokamakError2020, pharrErrorFieldPredictability2024,burschImprovedN1Empirical2026,luniaPredictedThresholdsRMP2025}. The metric most commonly used is known as the dominant mode overlap, $\delta$. This is a dimensionless, coordinate system-invariant quantity that follows from equation \eqref{eq:coupling_svd_trans2}. It quantifies the ability of an external flux to drive the tearing at the rational surfaces. In previous literature, $\delta$ has often been introduced without a fully explicit or consistent definition. This lack of precision obscures its interpretation, suggesting that $\delta$ varies between works. This is not the case. In this work, we provide a clear and consistent definition of $\delta$ that emphasizes its coordinate invariance. We also note that its implementation in the Generalized Perturbed Equilibrium Code (GPEC) \cite{parkIdealPerturbedEquilibria2009,parkGeneralizedPerturbedEquilibrium2018,wangModelingResistivePlasma2020} is fully compliant with the requirements for coordinate system invariance.


To avoid nonuniform weighting of rational surfaces and maintain the coordinate invariance of the LHS quantity, the coupling equation must be written using the quantities with magnetic field units in Eqs.~\ref{eq:b_half} and ~\ref{eq:b_full}, shown in Eq.~\ref{eq:coupling}. Decomposing this coupling matrix as before:
\begin{align}
    \bar{\mathbf{b}}^r &= U \, \Sigma \, V^\dagger \, \tilde{\mathbf{b}}^{x}, \label{eq:phir_svd}\\
    \bar{\mathbf{b}}^r_{(1)} &= \sigma_1 \, (\mathbf{v}_1^\dagger \tilde{\mathbf{b}}^x) \, \mathbf{u}_1, \label{eq:phir_dominant}
\end{align}
where the subscript $(1)$ denotes a rank-1 approximation. We find the approximate shielded field at each rational surface $\bar{\mathbf{b}}^r_{(1)}$, where $\bar{\mathbf{b}}^r_{(1)} \approx \bar{\mathbf{b}}^r$ when $\sigma_1 \gg \sigma_2$. From here the dominant mode overlap $\delta$ is defined as the inner product between $\tilde{\mathbf{b}}^x$ and the first right singular vector, $\mathbf{v}_1$, such that $\delta$ describes how much of the external magnetic field perturbation can effectively contribute to tearing at the rational surfaces. It is agnostic to the specifics of which rational surfaces are impacted, hence $\mathbf{u}_1$ is neglected, and by what factor, hence $\sigma_1$ is neglected. Therefore, the expression is correctly written as
\begin{equation}
    \delta = \frac{|\mathbf{v}_1^\dagger \tilde{\mathbf{b}}^x|}{B_T}.
    \label{eq:delta_dot_prod}
\end{equation}
This definition of $\delta$ is coordinate invariant, and is normalized by the toroidal field $B_T$ to be dimensionless. Future works that use $\delta$ should use this definition to ensure that their results are robust to coordinate system choice and are properly normalized. 


\begin{figure*}
\includegraphics[width=\linewidth]{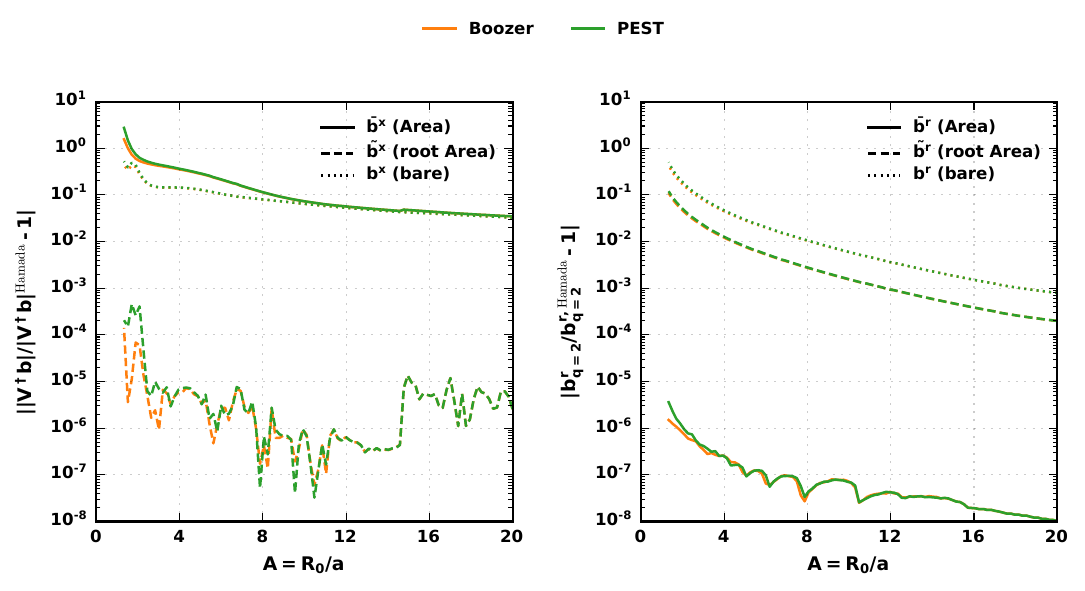}\caption{Aspect ratio and shaping scan plotted versus $A=R_0/a$ for Boozer (orange) and PEST (green) coordinates, showing relative deviation from Hamada coordinates. Line styles indicate the integrand weighting: solid for the area-weighted field ($\bar{\mathbf{b}}$), dashed for the root-area-weighted field ($\tilde{\mathbf{b}}$), and dotted for the bare, unweighted field ($\mathbf{b}$). \emph{Left:} the quadrature-sum overlap $|V^\dagger \mathbf{b}^x|$ defined in Eq.~\eqref{eq:overlap}, with the indicated weighting applied self-consistently to both $V$ and $\mathbf{b}^x$. \emph{Right:} the $q=2$ Fourier coefficient of the effective resonant field $\mathbf{b}^r$ under the indicated weighting. Only $\tilde{\mathbf{b}}^x$ yields a coordinate-invariant overlap (left) and only $\bar{\mathbf{b}}^r$ yields a coordinate-invariant resonant field (right). \label{fig:scan}}
\end{figure*}

\begin{figure*}
\includegraphics[width=\linewidth]{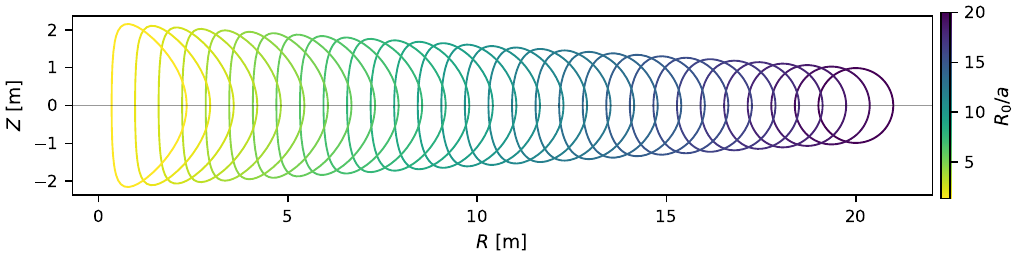}\caption{Last closed flux surface of equilibria used in the aspect ratio and shaping scan, shown in Fig.~\ref{fig:scan}. Scan ranges from inverse aspect ratio $\epsilon=0.1$ and circular cross section for cylinder-like physics to $\epsilon=0.75$, $\kappa=2.2$, $\delta=0.6$ for an extreme NSTX-U-like configuration.\label{fig:scanequils}}
\end{figure*}

The danger of using the wrong normalization is demonstrated with a shaping and aspect ratio scan in Fig.~\ref{fig:scan} using equilibria shown in Fig.~\ref{fig:scanequils}. The scan ranges from a near-circular, large aspect ratio equilibrium ($\epsilon=0.1$), representative of cylinder-like physics, to a strongly shaped, low aspect ratio, NSTX-U-like configuration ($\epsilon=0.75$, $\kappa=2.2$, $\delta=0.6$); the equilibria are generated with TokaMaker\cite{hansenOpenFUSIONToolkitOpenFUSIONToolkitV100beta52025,hansenTokaMakerOpensourceTimedependent2024}. Throughout the scan, a constant surface-normal perturbation of $10^{-4}\,\mathrm{T}$ is applied to the Hamada $m=2,3,4$ Fourier harmonics. The $q$ profile is allowed to vary but its ends remain fixed at $q_0=1.2$, $q_a=6.5$ by varying $I_P$ to keep $q=2,3,4,5,6$ in the computational domain. We choose to gauge relative invariance of computed quantities by coordinate systems with the quantity

\begin{equation}
    \left|V^\dagger \mathbf{b}^x\right| = \sqrt{\sum_i \left|V_i^\dagger \mathbf{b}^x\right|^2}, \label{eq:overlap}
\end{equation}

\noindent Here $V$ is the matrix of right singular vectors of the coupling matrix $C$, so $|V^\dagger \mathbf{b}^x|$ is the quadrature sum of the projections of the external field perturbation onto the dominant and subdominant modes (i.e. the dominant and subdominant mode overlaps), and is a measure of how much of the vacuum field perturbation can effectively couple to the plasma. The subdominant modes are included because the ordering of the singular values can change with coordinate system and with shaping, so the quadrature sum is a more robust diagnostic of coordinate dependence than the dominant overlap alone. This quantity is constructed self-consistently for each weighting: the same weighted field used to build $C$ (and hence $V$) is also the one being projected. The left panel of Fig.~\ref{fig:scan} therefore reports, for each line style, the divergence from Hamada of $|V^\dagger \mathbf{b}^x|$ where both $V$ and $\mathbf{b}^x$ correspond to the weighting indicated in the legend.

As shaping and inverse aspect ratio are increased, the coordinate dependence of the plasma-3D-field coupling for the different weightings of the vacuum field perturbation diverges significantly. The sum of the overlaps calculated with $\tilde{\mathbf{b}}^x$ is invariant across coordinate systems, remaining at a relative noise level of $\sim 10^{-5}$, while the sums calculated with $\mathbf{b}^x$ and $\bar{\mathbf{b}}^x$ diverge between coordinate systems by up to a factor of $\sim 2$--$3$ for the most extreme case.

The right panel of Fig.~\ref{fig:scan} shows the complementary quantity for the resonant interior: the coordinate divergence of the $q=2$ Fourier coefficient of the resonant field itself, evaluated under each weighting. This isolates the coordinate behavior of the physical resonant amplitude — the quantity that drives tearing — rather than its projection onto a singular basis. Only the full-area-weighted field $\bar{\mathbf{b}}^r$ is coordinate invariant; the bare and root-area-weighted resonant fields $\mathbf{b}^r$ and $\tilde{\mathbf{b}}^r$ exhibit strong coordinate dependence at low aspect ratio. While unshaped, large aspect ratio cases may only cause a 1-10\% variation in the resonant field and the vacuum field coupling, shapes and aspect ratios seen commonly in modern tokamaks can cause large discrepancies in both the coupling and the resonant field when not properly weighted. Studies using GPEC are robust by construction; the implementation has always paired the square-root-area weighted vacuum field $\tilde{\mathbf{b}}^x$ with the full-area-weighted resonant field $\bar{\mathbf{b}}^r$. The coordinate sensitivity instead (i) afflicts the three-mode metric (also known as TMEI), still in active use across work on multiple machines --- including DTT\cite{albaneseErrorFieldCorrection2023,Tesi_definitiva_Pasquale_Zumbolo,zumboloOptimalMagnetAssembly2024,iaiunesePreliminaryFaultAnalysis2025}, JT-60SA\cite{koNumericalStudyNovel2025,matsunagaInvesselCoilsMagnetic2015}, ITER coil and magnet-assembly design\cite{knasterITERNonaxisymmetricError2011,formisanoErrorFieldCorrection2010,olivaAnalysisErrorField2012,olivaEstimationErrorFields2013}, J-TEXT\cite{yangErrorFieldAnalysis2016}, and JA DEMO\cite{utohEstimationMagneticError2020} --- despite earlier critiques of its coordinate dependence\cite{parkMDC19ReportAssessment2017,amoskovAssessment1Overlap2019}; (ii) weakens ad-hoc resonant metrics defined without explicit attention to integrand weighting\cite{sugiyamaApplicationNonintegerRMP2026}; and (iii) is important to get right in emerging implementations of the coupling-matrix formalism in other codes such as MARS-F\cite{pigattoModelingdrivenRequirementsError2026a, mavigliaStudiesEUDEMO3D2023,liuModellingIntrinsicError2014}, which should ensure they appropriately adopt the weighting prescription given here. For example, work on MAST\cite{liuModellingIntrinsicError2014} found that the SVD-based optimization criterion did not match experimental EFC optima as well as direct cancellation of the $2/1$ resonant component; given the particular importance of the weighting in low aspect ratio, and in spite of the fact that this work uses a different singular coupling formalism, the necessity of the square-root-area weighting still applies and could contribute to the reported underperformance of the SVD approach. We note that these results use the SVD of a plasma response matrix that couples the vacuum field perturbation to the total field perturbation on the control surface, which is different than our coupling matrix $C$; in this case, both the domain and codomain of the coupling matrix should be weighted with the square-root-area factor to ensure coordinate invariance of the SVD and its modes. 

\section{\label{sec:discussion} Discussion}

This paper demonstrates that the SVD of the coupling matrix defined by $\bar{\mathbf{b}}^r = C \tilde{\mathbf{b}}^x$ is invariant under coordinate transformations among multiple systems of magnetic coordinates. This invariance holds when the resonant field on rational surfaces $\bar{\mathbf{b}}^r$ is defined according to the Park 2008\cite{parkSpectralAsymmetryDue2008} prescription with an area weighting in the Fourier decomposition, and the vacuum field perturbation $\tilde{\mathbf{b}}^x$ is weighted with a square-root-area factor. This means the real-space representations of the right singular vectors of $C$ are also invariant, and therefore the dominant mode is a physically meaningful quantity when $C$ is defined with these weightings. In contrast, the SVD of coupling matrices defined with other weightings of the vacuum field are coordinate-dependent, and their dominant modes are not physical. Park 2008 constrained the definition of the codomain of the coupling matrix by showing that the resonant Fourier coefficient on each rational surface is invariant under coordinate transformations when the Fourier integrand is weighted with the full area element. This paper constrains the definition of the domain of the coupling matrix in a complimentary fashion. The constraint presented in this paper has an even larger impact on the discrepancies that arise in its absence than the constraint of Park 2008; this is due to the differences between coordinate systems being more extreme at the plasma edge than in the plasma core as seen in Fig.~\ref{fig:coords}.

This result affirms the use of the existing GPEC paradigm for quantifying plasma-3D-field coupling \cite{parkErrorFieldCorrection2011, loganEmpiricalScaling22020, yangLocalizingResonantMagnetic2020}, and provides a clear prescription for how to construct coordinate-invariant coupling matrices and dominant modes. It also quantifies the coordinate dependence of coupling matrices and dominant modes constructed with other weightings of the vacuum field, and shows that this dependence can be significant for shaped, low aspect ratio equilibria. The coordinate dependence of the coupling matrix leads to a $\sim 1$--$10$\% deviation in the combined dominant and subdominant mode overlaps in high aspect ratio, cylindrical-like cases, but can lead to a factor of $\sim 2$--$3$ deviation in NSTX-U-like geometries, rendering the physical interpretation of results that use the wrong weighting of the vacuum field coordinate-dependent and potentially ambiguous.

GPEC has always used the correct weighting and its numerical results are robust to coordinate choice. However, notation in some published GPEC-based works has been ambiguous regarding which weighting is applied on which side of the coupling equation, which has obscured the requirement of coordinate invariance and may mislead future users. Future works should report the integrand weighting alongside any resonant or coupling result. The results of this paper are general and can be applied to any physical resonant metric defined by a Fourier spectrum, such as the effective resonant/shielded flux, the resonant field itself, or even non-field quantities such as $\Delta'$.

With both the domain and co-domain of $C$ constrained, the resonant coupling paradigm provides a coordinate-invariant basis for RMP coil design, error-field correction, and any other metric using Fourier decomposition to isolate resonant components on rational surfaces.

\begin{acknowledgments}
This work was supported by US DOE awards DE-SC0022270, DE-SC0022272, and DE-SC0024898. This work was also supported by the National Research Foundation of Korea (NRF), Ministry of Science and ICT, No. RS2023-00281276 and by R\&D Program of “Optimal Basic Design of DEMO Fusion Reactor, CN2502-1” through the Korea Institute of Fusion Energy(KFE) funded by the Government funds.
Data analysis and modeling in this work was facilitated by the OMFIT integrated modeling framework\cite{meneghiniIntegratedModelingApplications2015}. 
\end{acknowledgments}

\section*{Data Availability Statement}

The simulation data that support the findings of this study are available from the corresponding author upon reasonable request.

\bibliography{references}

\end{document}